\documentclass[sort&compress
    ,final            
  ,numberedheadings 
  ]
  {aipproc}

\layoutstyle{8x11single}
   \usepackage{showkeys}
	\usepackage{amssymb}
	\usepackage{amsmath}

\newcommand\beq{\begin{equation}}
\newcommand\eeq{\end{equation}}
\newcommand\beqa{\begin{eqnarray}}
\newcommand\eeqa{\end{eqnarray}}
\newcommand{\nn}{\nonumber\\}

\begin{document}

\title{Kinetic Theory of Soft Matter. The Penetrable-Sphere Model}

\author{Andr\'es Santos}{
  address={Departamento de F\'{\i}sica, Universidad de Extremadura, E-06071 Badajoz, Spain}
}



\begin{abstract}
 The penetrable-sphere model has been introduced in the literature to describe the peculiar thermodynamic behavior of some colloidal systems. In this model the interaction potential is $\phi(r)=\epsilon>0$ if the two spheres are overlapped ($r<\sigma$) and $\phi(r)=0$ otherwise ($r>\sigma$). In this paper the shear viscosity, thermal conductivity, and self-diffusion coefficients of a dilute gas of penetrable spheres are evaluated.  It is found that the effective collision frequency $\nu(T^*)$ grows as $\sqrt{T^*}$  up to $T^*\equiv k_BT/\epsilon\simeq 0.25$, reaches a maximum at $T^*\simeq 0.415$ and then  decays as ${T^*}^{-3/2}\log T^*$ for large temperatures. The results are applied to the hydrodynamic profiles in the steady Fourier and Couette flows.
\end{abstract}

\maketitle


\section{Introduction}
Traditionally, the kinetic theory of gases has been applied to gases made of particles which interact according to \textit{unbounded} spherically symmetric pair potentials, such as hard spheres, power-law repulsive interactions, the square-well model, or the Lennard--Jones potential  \cite{CC70}. Moreover, granular gases are usually modeled as constituted by inelastic hard spheres. In the last decade, however, the equilibrium properties of fluids of particles interacting via \textit{bounded} pair potentials have been the subject of increasing interest, the Gaussian core model and the penetrable-sphere model being among the most popular ones \cite{L01,AS04}. These models have been proposed in the literature in order to understand the peculiar behavior of some \textit{colloidal} systems, such as micelles in a solvent or star copolymer suspensions. The particles in these colloids are constituted by a small core surrounded by several attached polymeric arms. As a consequence of their structure, two or more of these particles allow a considerable degree of overlapping with a small energy cost \cite{L01}. These are a few particular cases of systems defining what is commonly known as ``soft matter,'' which has become an active field of research with interesting physical, chemical, and engineering applications. 

While the equilibrium properties and phase diagrams of particles interacting through  soft potentials have been studied with great detail,  the nonequilibrium transport properties of those systems seem to have received less attention. The aim of this paper is to contribute to the understanding of the nonequilibrium properties of particles interacting via bounded potentials by analyzing the transport coefficients of the penetrable-sphere (PS) model in the low-density regime. 

\section{Collision process in the penetrable-sphere model}
The PS interaction potential is defined as 
\beq
\phi(r)=\left\{
\begin{array}{ll}
\epsilon>0,&r<\sigma,\\
0,&r>\sigma.
\end{array}
\right.
\label{n1}
\eeq
 Therefore, in this model the gas behaves as a hard-sphere (HS) gas in the low-temperature limit ($T^*\equiv k_BT/\epsilon\to 0$) and as a collisionless gas in the high-temperature limit ($T^*\to\infty$). 
The Liouville operator and the Boltzmann--Lorentz collision operator for this model have been derived in Refs.\ \cite{A75} and \cite{GDL79}, respectively.
The PS model should not be confused with the Widom--Rowlinson (WR) model of interpenetrating spheres \cite{WR70}, which is thermodynamically equivalent to a binary mixture of highly non-additive hard spheres. The transport properties of the WR model are obtained in Ref.\ \cite{KS79}.

\begin{figure}[tbp]
\includegraphics[width=0.81\columnwidth]{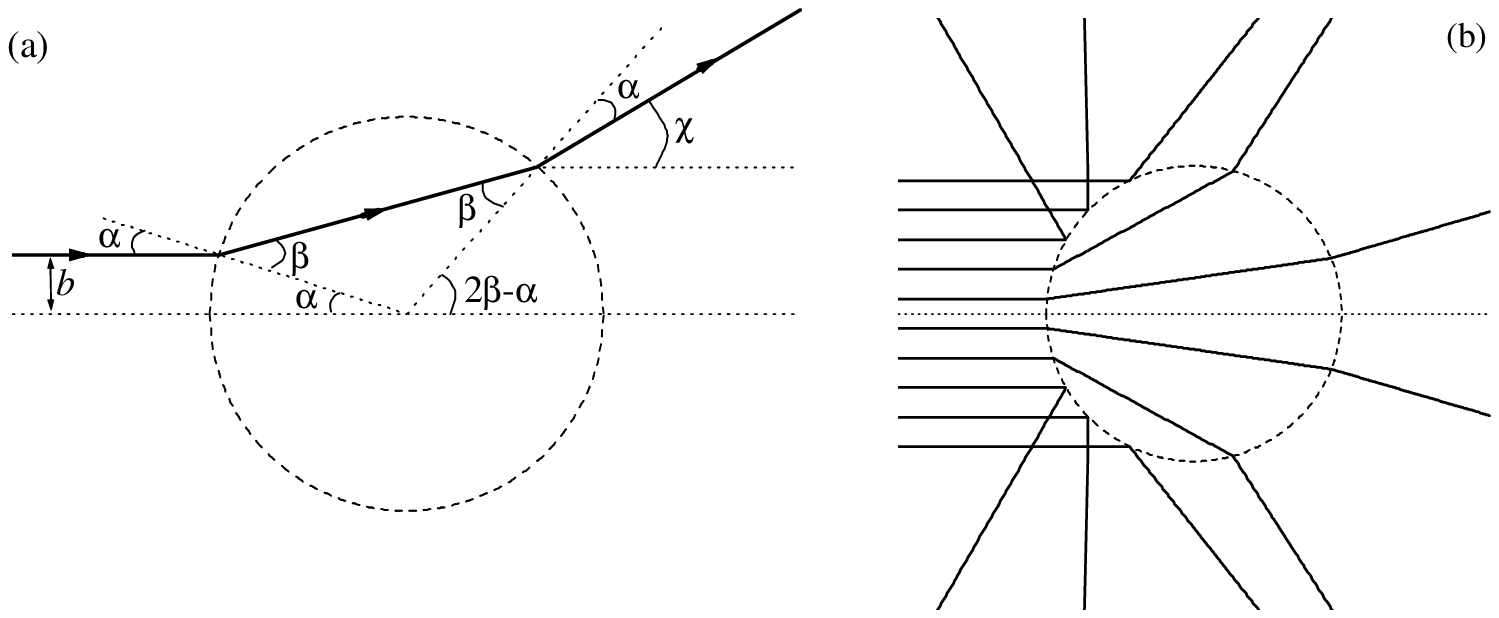}
\caption{(a) Sketch of the collision process when $g^*>1$ and $b^*<n(g^*)$. The incoming incidence and refraction angles are $\alpha=\sin^{-1}b^*$  and $\beta=\sin^{-1}\left[b^*/n(g^*)\right]$, respectively, while the outgoing angles are $\beta$ and $\alpha$, respectively. Thus, the scattering angle is $\chi=2(\beta-\alpha)$. This sketch actually corresponds to $g^*=1.2$ and $b^*=0.3$. (b) Different possible trajectories for $g^*=1.1$.
\label{sketch}}
\end{figure}

\begin{figure}[tbp]
\includegraphics[width=1.0\columnwidth]{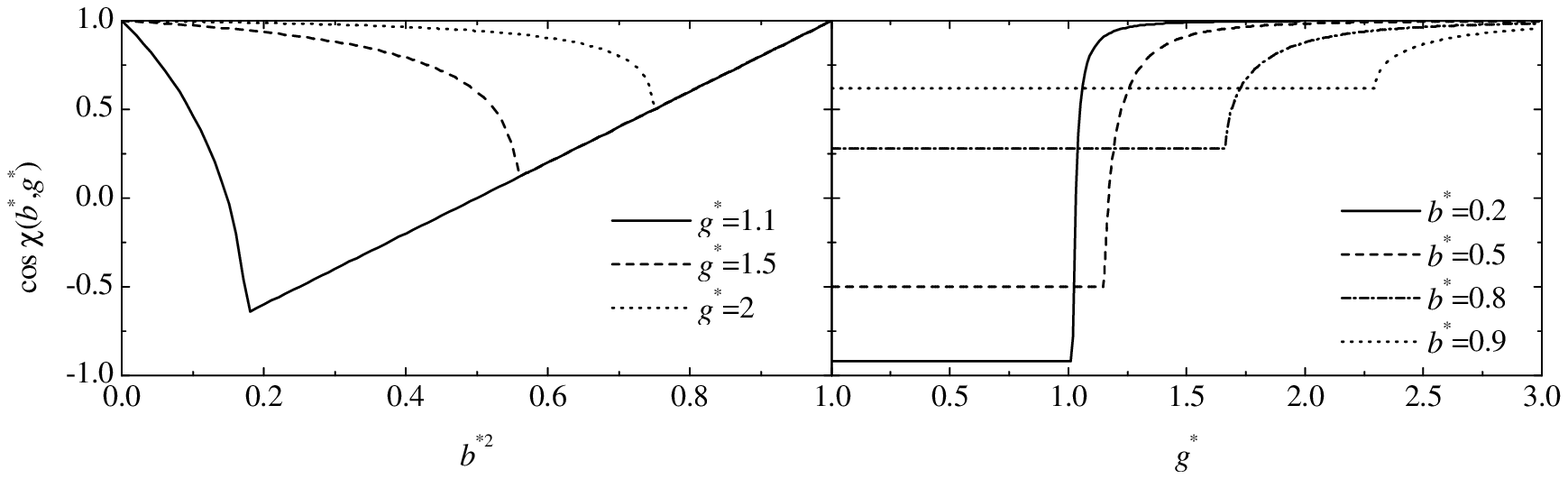}
\caption{Left panel: Plot of $\cos\chi(b^*,g^*)$ versus ${b^*}^2$ for  $g^*=1.1$, $g^*=1.5$, and $g^*=2$. Right panel: $\cos\chi(b^*,g^*)$ versus ${g^*}$ for $b^*=0.2$, $b^*=0.5$, $b^*=0.8$, and $b^*=0.9$.
\label{coschi}}
\end{figure}

Before delving into the transport coefficients of the PS model, let us analyze its peculiar binary collision process.
As usual, we consider the equivalent one-body problem in which a projectile particle (with the reduced mass $\mu=m/2$ 
 of the colliding pair) feels the central potential $\phi(r)$ centered at the origin. The projectile  approaches the ``target'' with a (relative) speed $g$ and an impact parameter $b$, being  deflected after interaction with a scattering angle $\chi(b^*,g^*)$ that depends on the reduced impact parameter $b^*\equiv b/\sigma$ and reduced speed $g^*\equiv g/\sqrt{2\epsilon/\mu}$. Obviously, if $g^*<1$, the projectile does not have enough kinetic energy to ``penetrate'' into the core of the target and consequently it is deflected exactly as if the target were a hard sphere; the associated scattering angle is therefore  $\chi(b^*,g^*)={\pi}-2\sin^{-1} b^*$. On the other hand, if $g^*> 1$, the projectile traverses the core of the target moving with a (reduced) kinetic energy ${g^*}^2-1$ and eventually leaves the core along the scattering direction with the same speed $g^*$ as before collision. This penetration process is analogous to the double refraction of light through a sphere made of a transparent material of relative refraction index  $n(g^*)\equiv\sqrt{1-{g^*}^{-2}}<1$. Figure \ref{sketch}(a) shows a sketch of a typical collision with $g^*>1$. 
The scattering angle in this case is $\chi(b^*,g^*)=2\sin^{-1}\left[{b^*}/n(g^*)\right]-2\sin^{-1} b^*$. However, the latter expression is valid only if the impact parameter is sufficiently small, namely if ${b^*}<n(g^*)$. Otherwise, the incidence angle $\alpha=\sin^{-1}b^*$ is large enough as to produce a ``total reflection'' effect, so the projectile is again deflected as if colliding with a hard sphere [cf.\ Fig.\ \ref{sketch}(b)]. In summary, the scattering angle $\chi(b^*,g^*)$ for the PS potential is given by
\beq
\cos\chi(b^*,g^*)=2{b^*}^2-1+\left\{
\begin{array}{ll}
\Delta_1(b^*,g^*),& g^*>1 \text{ and } 0\leq b^*\leq n(g^*),\\
0,& \text{otherwise}.
\end{array}
\right.
\label{4}
\eeq
where
\beq
\Delta_1(b^*,g^*)=2\sqrt{1-{b^*}^2/n^2(g^*)}\left[2 \left({b^*}^2/n(g^*)\right)\sqrt{1-{b^*}^2}-\left(2{b^*}^2-1\right)\sqrt{1-{b^*}^2/n^2(g^*)}\right].
\label{5}
\eeq
 At a given value of the (reduced) relative speed $g^*>1$, $\chi$ starts increasing with $b^*$ until it reaches a maximum value 
$ \chi_{\text{max}}(g^*)=\cos^{-1}\left(1-2 {g^*}^{-2}\right)$
at 
$b^*=n(g^*)$, and  decreases thereafter. Figure \ref{coschi} shows $\cos\chi$ for a few cases.

\begin{figure}[tbp]
\includegraphics[width=.65\columnwidth]{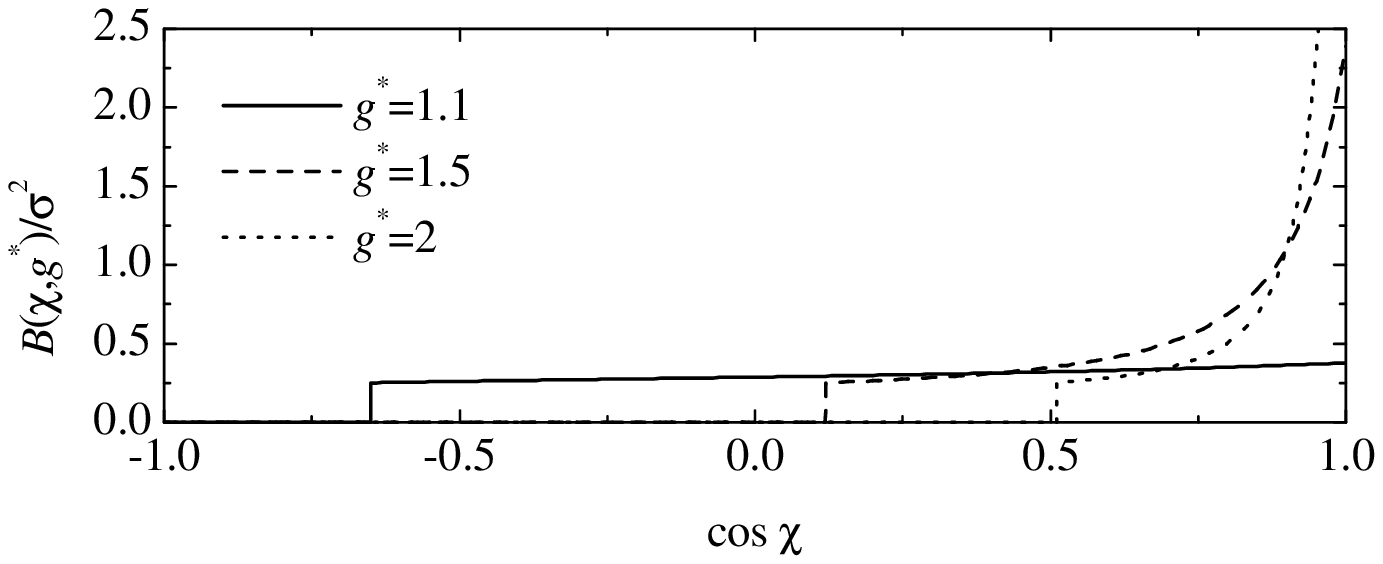}
\caption{Differential cross section $B(\chi,g^*)/\sigma^2$ versus $\cos\chi$ for $g^*=1.1$, $g^*=1.5$, and $g^*=2$.
\label{cross}}
\end{figure}

{}From the scattering law one can obtain the \textit{differential} cross section as $B(\chi,g^*)=\frac{1}{2}\sigma^2|\partial {b^*}^2/\partial\cos\chi|$. If $g^*<1$, then $B(\chi,g^*)=\frac{1}{4}\sigma^2$, as for hard spheres. Otherwise,
\beq
B(\chi,g^*)/\sigma^2=\left\{
\begin{array}{ll}
0, &\chi>\chi_{\text{max}}({g^*}),\\
\frac{1}{4}+\frac{1}{2}\left|\frac{\partial {b^*}^2}{\partial\cos\chi}\right|_{{b^*}<n(g^*)},&
0\leq \chi<\chi_{\text{max}}({g^*}),
\end{array}
\right.
\label{5.1}
\eeq
where
\beq
\left|\frac{\partial {b^*}^2}{\partial\cos\chi}\right|_{{b^*}<n(g^*)}=
({g^*}^2-1)\frac{2{\sqrt{1+\cos\chi}} (2{g^*}^2-1)+ 
{g^*}{{\sqrt{{2({g^*}^2}-1)}}} \left[1+3 \cos\chi-2 {{\left(1-\cos\chi\right)}^2} {{g^*}^2}({g^*}^2-1)\right]}{ 4{\sqrt{1+\cos\chi}}{{\left[1+2(1-\cos\chi){{g^*}^2}({{g^*}^2}-1)\right]}^2}}.
\label{5.2}
\eeq
Figure \ref{cross} shows the differential cross section $B(\chi,g^*)$ for several values of $g^*>1$. We observe that $B(\chi,g^*)=0$ for $\chi>\chi_{\text{max}}(g^*)$, i.e. for $\cos\chi<1-2{g^*}^{-2}$, takes the HS value $B(\chi,g^*)=\frac{1}{4}\sigma^2$ at $\chi=\chi_{\text{max}}^-(g^*)$ and monotonically increases with decreasing scattering angle for  $\chi<\chi_{\text{max}}(g^*)$, i.e. for $\cos\chi>1-2{g^*}^{-2}$. This implies that, as expected, when the kinetic energy is much larger than $\epsilon$ most of the collisions are grazing. On the other hand, the \textit{total} cross section, which is given by the area below the curves in Fig.\ \ref{cross}, is independent of $g^*$ and hence it coincides with that of HS, namely 
$2\pi\int_{1-2/{g^*}^{2}}^1 d(\cos\chi)\,  B(\chi,g^*)=\pi\sigma^2$.

\section{Transport coefficients}
As is well known, the Chapman--Enskog method allows one to derive the Navier--Stokes transport coefficients from the Boltzmann equation for  a dilute gas   in terms of the scattering law corresponding to the interaction potential of interest \cite{CC70}. For a single gas, the relevant transport coefficients are the shear viscosity $\eta(T)$, the thermal conductivity $\kappa(T)$, and the self-diffusion coefficient $D(T)$. In the first Sonine approximation, their expressions are \cite{CC70}
\beq
\eta(T)=\frac{5}{8}\frac{k_BT}{\Omega_{2,2}(T)},\quad \kappa(T)=\frac{15}{4}\frac{k_B}{m}\eta(T),\quad D(T)=\frac{3}{8}\frac{k_BT}{mn\Omega_{1,1}(T)},
\label{6}
\eeq
where the collisional integrals $\Omega_{k,\ell}(T)$ are
\beq
\Omega_{k,\ell}(T)=\sqrt{\frac{k_BT}{2\pi\mu}}\int_0^\infty dy\,e^{-y^2}y^{2k+3}Q_{\ell}\left(y\sqrt{2k_BT/\mu}\right),\quad Q_\ell(g)=2\pi \int_0^\infty db\, b\left[1-\cos^\ell\chi(b,g)\right].
\label{7}
\eeq
In the special case of hard spheres, one has
\beq
\Omega_{k,\ell}^{\text{HS}}(T)=\pi\sigma^2\frac{(k+1)!}{2}\frac{1-\frac{1+(-1)^\ell}{2(1+\ell)}}{\sqrt{2\pi \mu/k_BT}},
\label{n3}
\eeq
\beq
\eta_{\text{HS}}(T)=\frac{5}{16}\frac{\sqrt{mk_BT/\pi}}{\sigma^2},\quad \kappa_{\text{HS}}(T)=\frac{15}{4}\frac{k_B}{m}\eta_{\text{HS}}(T),\quad D_{\text{HS}}(T)=\frac{3}{8}\frac{\sqrt{k_BT/m\pi}}{n\sigma^2}.
\label{n2}
\eeq

Now we particularize to penetrable spheres. In that case, it is convenient to define the 
\textit{reduced} integrals $\Omega_{k,\ell}^*(T^*)=\Omega_{k,\ell}(T)/\Omega_{k,\ell}^{\text{HS}}(T)$, so that
\beq
{\eta_{\text{HS}}}/\eta={\Omega_{2,2}^*(T^*)}, \quad 
{\kappa_{\text{HS}}}/\kappa={\Omega_{2,2}^*(T^*)}, \quad 
{D_{\text{HS}}}/D={\Omega_{1,1}^*(T^*)}.
\label{9}
\eeq
The functions $\Omega_{k,\ell}^*(T^*)$ can be expressed as
\beq
\Omega_{k,\ell}^*(T^*)
=1-\frac{4}{{(k+1)!}\left[1-\frac{1+(-1)^\ell}{2(1+\ell)}\right]}\int_{1/\sqrt{T^*}}^\infty dy\,
e^{-y^2}y^{2k+3}R_{\ell}\left(y\sqrt{T^*}\right),
\label{10}
\eeq
where we have called
\beq
R_\ell(g^*)=\int_0^{n(g^*)}db^*\, b^* \Delta_\ell(b^*,g^*),\quad
\Delta_\ell(b^*,g^*)\equiv \cos^\ell\chi(b^*,g^*)-\left(2{b^*}^2-1\right)^\ell.
\label{11}
\eeq
In particular, taking into account Eq.\ (\ref{4}) and setting $\ell=1$ and $\ell=2$, one gets
\beqa
R_1(g^*)&=&\frac{1}{6{g^*}^4}\left({g^*}^2-1\right)\left({g^*}^2+2\right)+\frac{1}{12{g^*}^3}\left({g^*}^2-1\right)^{-1/2}\left(4{g^*}^4-4{g^*}^2+3\right)\nn
&&
+\frac{1}{8{g^*}^4}\left({g^*}^2-1\right)^{-1}\left(2{g^*}^2-1\right)\ln\left(2{g^*}^2-2{g^*}\sqrt{{g^*}^2-1}-1\right),
\label{12}
\eeqa
\beqa
R_2(g^*)&=&
\frac{1}{15{g^*}^6}\left({g^*}^2-1\right)\left(3{g^*}^4+4{g^*}^2-12\right)+  \frac{1}{60{g^*}^5}({g^*}^2-1)^{-3/2}\left(8{g^*}^8-16{g^*}^6+58{g^*}^4-50{g^*}^2+15\right) \nn
&& +\frac{1}{8{g^*}^6}\left({g^*}^2-1\right)^{-2}\left(2{g^*}^2-1\right) \left(2{g^*}^4-2{g^*}^2+1\right)
\ln\left(2{g^*}^2-2{g^*}\sqrt{{g^*}^2-1}-1\right).
\label{13}
\eeqa

\begin{figure}[tbp]
\includegraphics[width=1.0\columnwidth]{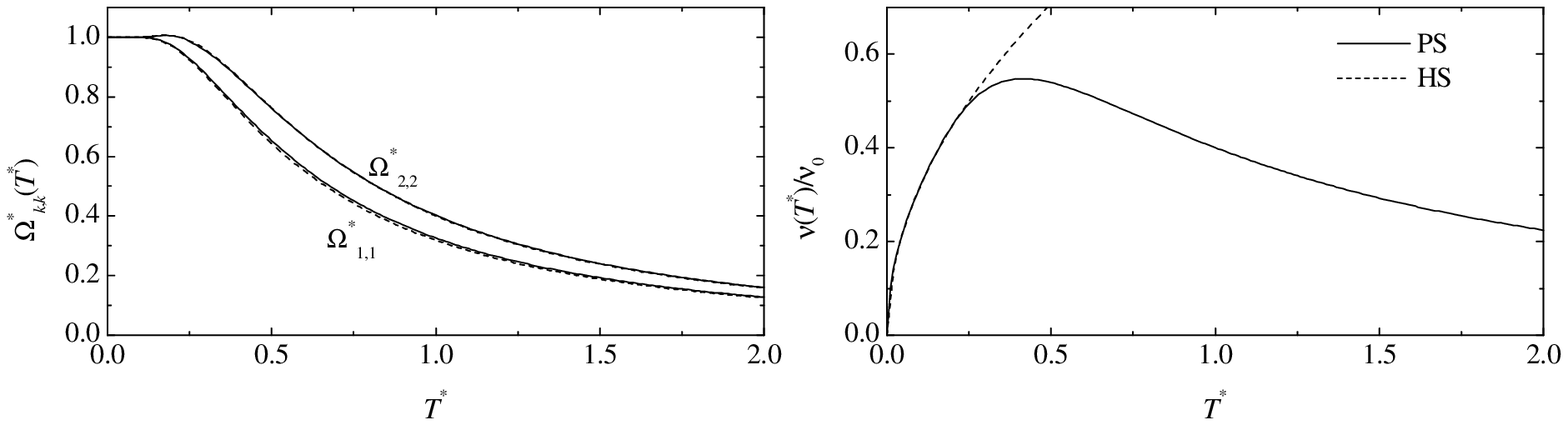}
\caption{Left panel: $\Omega_{1,1}^*(T^*)$ and $\Omega_{2,2}^*(T^*)$ as obtained from Eq.\ (\protect\ref{10}) by numerical integration (solid lines) and from the approximate expressions (\protect\ref{16}) and (\protect\ref{17}) (dashed lines). The numerical and approximate curves are practically indistinguishable in the case of $\Omega_{2,2}^*(T^*)$. Right panel: Effective collision frequency $\nu(T^*)/\nu_0$ for the PS model (solid line) and for HS (dashed line).
\label{omega}}
\end{figure}

Insertion of Eqs.\ (\ref{12}) and (\ref{13}) into Eq.\ (\ref{10}) gives $\Omega_{1,1}^*(T^*)$ and $\Omega_{2,2}^*(T^*)$ in terms of quadratures that need to be evaluated numerically.
On the other hand, for high temperatures ($T^*\gg 1$) the integral in Eq.\ (\ref{10}) is dominated by the behavior of $R_\ell(g^*)$  for $g^*\gg 1$. In that limit Eqs.\ (\ref{12}) and (\ref{13}) become 
\beq
R_1(g^*)=\frac{1}{2}-\frac{1+4\ln\left(2g^*\right)}{8{g^*}^4}+{O}\left({g^*}^{-6}\ln g^*\right)
,\quad 
R_2(g^*)=\frac{1}{3}-\frac{1+4\ln\left(2g^*\right)}{4{g^*}^4}+{O}\left({g^*}^{-6}\ln g^*\right).
\label{15}
\eeq
This yields the following \textit{asymptotic} approximations: 
\beq
\Omega_{1,1}^*(T^*)\approx 1-e^{-1/T^*}\left(1+\frac{1}{T^*}\right)-\frac{\text{Ei}(-1/T^*)}{4{T^*}^2},
\label{16}
\eeq
\beq
\Omega_{2,2}^*(T^*)\approx 1-e^{-1/T^*}\left(1+\frac{1}{T^*}-\frac{4\ln 2-1}{8{T^*}^2}\right)-\frac{\text{Ei}(-1/T^*)}{4{T^*}^2},
\label{17}
\eeq
where 
$\text{Ei}(z)=-\Gamma(0,-z)=-\int_{-z}^\infty dt\, t^{-1}{e^{-t}}$
is the exponential integral function, $\Gamma(n,z)$ being the incomplete gamma function.
Quite interestingly, the approximate expressions (\ref{16}) and (\ref{17}) agree almost perfectly for the whole temperature domain with the results obtained from Eq.\ (\ref{10}) by numerical integration, as shown in the left panel of Fig.\ \ref{omega}. As usual, we can introduce an \textit{effective} collision frequency associated with the shear viscosity as \cite{CC70}  $\nu(T)=nk_BT/\eta(T)$. Therefore,
\beq
\nu(T^*)=\nu_0\sqrt{T^*}\Omega_{2,2}^*(T^*),\quad \nu_0\equiv \frac{16}{5}n\sigma^2\sqrt{\pi\epsilon/m}.
\label{19}
\eeq
In the case of hard spheres, $\nu_{\text{HS}}(T^*)=\nu_0 \sqrt{T^*}$. The right panel of Fig.\ \ref{omega}  shows $\nu(T^*)/\nu_0$ for PS and HS. We observe that up to $T^*\simeq 0.25$ both systems have practically the same collision frequency. For larger temperatures, however, both collision frequencies strongly differ each other. While in the HS case the mean free path is independent of temperature and hence $\nu_{\text{HS}}(T^*)$ grows proportionally to the thermal velocity,   penetrability effects in the PS model become more and more important for $T^*\gtrsim 0.25$. As a consequence, $\nu(T^*)$ reaches a maximum value $\nu_{\text{max}}\simeq 0.548 \nu_0$ at $T^*=T^*_{\text{max}}\simeq 0.415$, and then decays for asymptotically large temperatures as $\nu(T^*)\approx {T^*}^{-3/2}\left(\frac{1}{4}\ln T^*+\frac{3}{8}-\frac{1}{4}\gamma+\ln\sqrt{2}\right)$, where $\gamma\simeq 0.577216$ is Euler's constant.
\section{Application to the Fourier and Couette flows}
Let us compare now the temperature and velocity profiles for HS and PS in the steady Fourier and Couette flows \cite{GS03}. In the planar Fourier flow the gas is enclosed between two parallel plates at rest located at $y=0$ and $y=L$ and kept at temperatures $T_1$ and $T_2$, respectively. In the steady state described by the Navier--Stokes equations the temperature profile is the solution to $\nu^{-1}(T)\partial T/\partial y=\mbox{const}$. Neglecting boundary layer effects and applying the boundary conditions, the implicit solution is
\beq
y/L=\left[{\int_{T_1^*}^{T_2^*}d\theta\,\nu^{-1}(\theta)}\right]^{-1}{\int_{T_1^*}^{T^*}d\theta\,\nu^{-1}(\theta)}.
\label{20}
\eeq
In the case of hard spheres, this yields $T^*(y)=T_1^*\left[1+\left(\sqrt{T_2^*/T_1^*}-1\right)y/L\right]^2$.

In the planar Couette flow the plates at $y=0$ and $y=L$ move along the $x$-direction with velocities $-U$ and $+U$, respectively,  but otherwise they are kept at the same temperature $T_0$. According to the Navier--Stokes equation in the steady state the temperature and velocity fields are related  \citep[pp.\ 213--222]{GS03} by $T^*(u_x^*)=T_0^*\left[1+M^2(1-{u_x^*}^2)\right]$, where $u_x^*=u_x/U$ is the reduced flow velocity and $M\equiv \sqrt{2mU^2/15k_BT_0}$ is a sort of Mach number. The velocity profile obeys the condition $\nu^{-1}(T(u_x))\partial u_x/\partial y=\mbox{const}$, whose implicit solution (again neglecting boundary effects) is
\beq
y/L=\left[{\int_{-1}^{1}dw\,\nu^{-1}(T^*(w))}\right]^{-1}{\int_{-1}^{u_x^*}dw\,\nu^{-1}(T^*(w))}.
\label{21}
\eeq
For hard spheres, 
$u_x^*(y)=\sqrt{1+M^{-2}}\sin\left[\left(2y/L-1\right)\sin^{-1}(1+M^{-2})^{-1/2}\right]$.
\begin{figure}[h!]
\includegraphics[width=1.0\columnwidth]{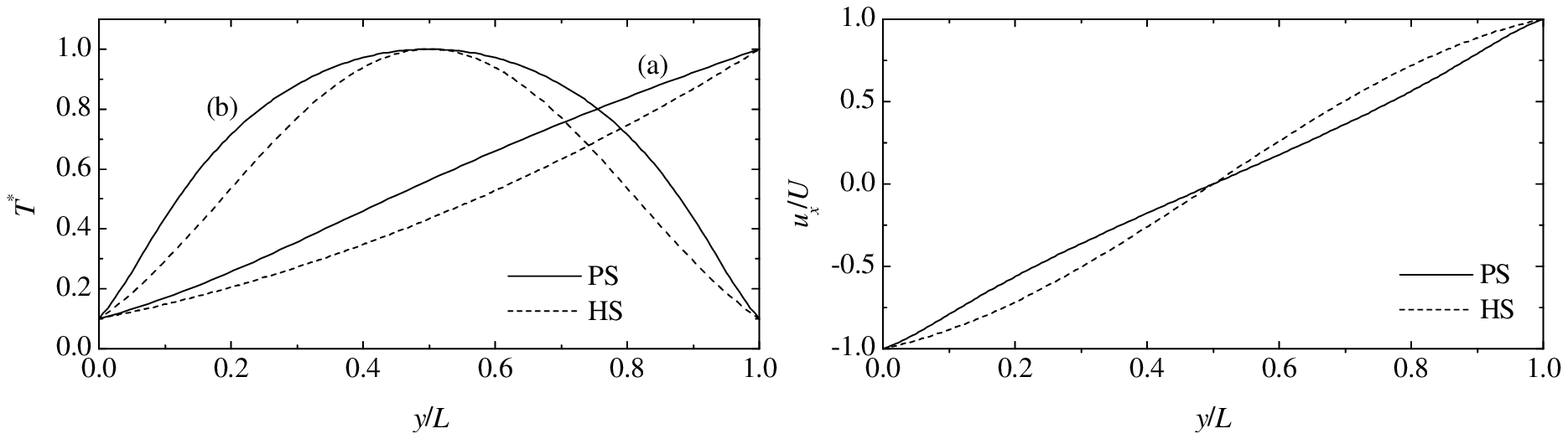}
\caption{Left panel: Temperature profiles for (a) the Fourier flow with $T_1^*=0.1$ and $T_2^*=1$ and (b) the Couette flow with $T_0=0.1$ and $M\equiv \sqrt{2mU^2/15k_BT_0}=3$. Right panel: Velocity profile for the Couette flow with $T_0=0.1$ and $M=3$. 
\label{profiles}}
\end{figure}

The temperature and velocity profiles corresponding to HS and PS are compared in Fig.\ \ref{profiles} for representative examples of Fourier and Couette flows. Note that in the region where $T^*>T^*_{\text{max}}\simeq 0.415$, the PS and HS systems have opposite curvatures for the temperature (Fourier) and velocity (Couette) profiles.

\section{Concluding remarks}
The aim of this paper has been to derive and analyze the Navier--Stokes transport coefficients of a dilute gas made of penetrable spheres (PS). Apart from its ability to describe qualitatively the effective interactions in some colloidal systems, the PS model is also interesting as a crossover between a gas of hard spheres (HS) in the low temperature limit ($T^*\to 0$) and a gas of non-interacting particles in the opposite limit ($T^*\to \infty$).
Although the total cross section of PS at any temperature is the same as that of HS, the effective collision frequency $\nu(T^*)$ associated with the shear viscosity and thermal conductivity exhibits a non-trivial temperature dependence. As expected on physical grounds, $\nu(T)$ ranges from $\nu(T^*)\to \nu_{\text{HS}}(T^*)\propto\sqrt{T^*}$ when $T^*\to 0$ to $\nu(T^*)\to 0$ when $T^*\to \infty$, so it  presents a maximum value $\nu_{\text{max}}$ at a certain temperature $T_{\text{max}}^*$. On the other hand, plausible physical arguments do not suffice to answer more specific questions, for instance: (i) at which threshold temperature $T_{\text{th}}^*$ does $\nu(T^*)$ start to deviate significantly from $\nu_{\text{HS}}(T^*)$?; (ii) what is the value of $T_{\text{max}}^*$?; (iii) how small is $\nu_{\text{max}}$ as compared with $\nu_{\text{HS}}(T_{\text{max}}^*)$?; (iv) at which temperature $T_{1/2}^*>T_{\text{max}}^*$ the collision frequency has decayed to half its maximum value?; (v) how is the asymptotic behavior of $\nu(T^*)$ in the high-temperature limit?

The evaluation of the transport coefficients and of $\nu(T^*)$ has required the detailed analysis of the collision process. In principle, the collision integrals must be evaluated numerically [cf.\ Eqs.\ (\ref{10})--(\ref{13})], but excellent approximations  are provided by Eqs.\ (\ref{16}) and (\ref{17}). The answer to the questions posed above are: (i) $T_{\text{th}}^*\simeq 0.25$; (ii) $T_{\text{max}}^*\simeq 0.415$; (iii) $\nu_{\text{max}}/\nu_{\text{HS}}(T_{\text{max}}^*)\simeq 0.850$; (iv) $T_{1/2}^*\simeq 1.619$; (v) $\nu(T^*)\approx \frac{1}{4}{T^*}^{-3/2}\ln T^*$ for $T^*\to\infty$. The contrasting temperature dependencies of the PS and HS transport coefficients for $T^*>T_{\text{th}}^*\simeq 0.25$ have been illustrated here by comparing the respective hydrodynamic profiles for the Fourier and Couette flows.


\begin{theacknowledgments}
It is a pleasure to dedicate this paper to Carlo Cercignani, who has taught us so much about kinetic theory.
The author is grateful to A. V. Bobylev and J. W. Dufty for helpful comments.
Partial support from the Ministerio de   
Educaci\'on y Ciencia 
(Spain) through grant No.\ FIS2004-01399 is acknowledged.  
\end{theacknowledgments}


\bibliographystyle{aipproc}   

\end{document}